\documentclass[10pt,letterpaper,english,reprint, aps]{revtex4-1}
\usepackage{mathptmx}
\usepackage{helvet}

\usepackage[LGR,T1]{fontenc}
\usepackage[latin9]{inputenc}
\usepackage{textcomp}
\usepackage{amsmath}
\usepackage{stackrel}
\usepackage{graphicx}
\usepackage{esint}
\usepackage{subscript}

\makeatletter

\pdfpageheight\paperheight
\pdfpagewidth\paperwidth

\DeclareRobustCommand{\greektext}{%
  \fontencoding{LGR}\selectfont\def\encodingdefault{LGR}}
\DeclareRobustCommand{\textgreek}[1]{\leavevmode{%
  \IfFileExists{grtm10.tfm}{}{\fontfamily{cmr}}\greektext #1}}
\DeclareFontEncoding{LGR}{}{}
\DeclareTextSymbol{\~}{LGR}{126}
\providecommand{\tabularnewline}{\\}

\makeatother

\usepackage{babel}
\graphicspath{./}
\begin{document}

\title{Fast calculation of two-electron-repulsion integrals: a numerical
approach}

\author{Pedro E. M. Lopes}

\affiliation{Rua Almirante Reis, Nº 28A, 2º Esq, 2330-099 Entroncamento, Portugal }

\altaffiliation{11 Warren Lodge Ct 1D, Cockeysville, MD 21030, USA}

\email{plopesuk@yahoo.co.uk }

\homepage{www.fastcompchem.com }

\selectlanguage{english}%
\begin{abstract}
An alternative methodology to evaluate two-electron-repulsion integrals
based on numerical approximation is proposed. Computational chemistry
has branched into two major fields with methodologies based on quantum
mechanics and classical force fields. However, there are significant
shadowy areas not covered by any of the available methods. Many relevant
systems are often too big for traditional quantum chemical methods
while being chemically too complex for classical force fields. Examples
include systems in nanomedicine, studies of metalloproteins, etc.
There is an urgent need to develop fast quantum chemical methods able
to study large and complex systems. This work is a proof-of-concept
on the numerical techniques required to develop accurate and computationally
efficient algorithms for the fast calculation of electron-repulsion
integrals, one of the most significant bottlenecks in the extension
of quantum chemistry to large systems. All concepts and calculations
were developed for the three-center integral $\left(p_{xA}p_{xB}|p_{xC}p_{xC}\right)$
with all atoms being carbon. Starting with the analytical formulae,
convenient decompositions were tested to provide smooth two-dimensional
surfaces that were easily fitted. The approximating algorithm consisted
of a multilayered approach based on multiple fittings of two-dimensional
surfaces. An important aspect of the new method is its independence
on the number of contracted Gaussian primitives. The basis set of
choice was STO-6G. In future developments, larger basis sets will
be developed. This work is part of a large effort aimed at improving
the inadequacies of existing computational chemistry methods, both
based on quantum mechanics and classical force fields, in particular
in describing large and heterogeneous systems (ex. metalloproteins).
\end{abstract}


\maketitle

\section{\label{sec:Introduction 1}Introduction }

The field of computational quantum chemistry has experienced extraordinary
progress to date due to advances in computing power and the development
of new algorithms. While advances have been reached, still there are
limitations in the size and/or complexity of the systems that can
be studied. In the second decade of the twenty-first century the words
of Paul Dirac in 1929 {[}1{]} still echo: \textit{\textquotedblleft the
underlying physical laws necessary for the mathematical theory of
a large part of physics and the whole of chemistry are thus completely
known, and the difficulty is only that the exact application of these
laws leads to equations much too complicated to be soluble.\textquotedblright{}}
Today, Dirac\textquoteright s statement still remains true and many
of the equations governing the chemical phenomena are still too complex
to solve using today\textquoteright s computational resources. 

To answer complex chemical phenomena, computational quantum chemistry
has suffered multiple numerical approximations and simplifications.
Some are numerical approximations to the fundamental equations such
as the Born-Oppenheimer approximation that greatly simplifies the
Schrödinger equation by considering that the much heavier nuclei remain
stationary during the calculation. Other approximations, leading for
example to empirical and semi-empirical methods, consider simplified
forms of the first-principles underlying equations that are typically
faster to solve. Other classes of methods in computational chemistry
have abandoned the quantum chemical principles altogether and classical
approximations to the potential energy surface based on force fields
were developed, as they are computationally less intensive than quantum
chemical electronic structure calculations. Empirical force fields
are currently the methods of choice for studies of large systems in
biology and materials science, for example conformational studies
of proteins, DNA, etc and protein-ligand binding thermodynamics. However,
empirical force fields have severe limitations: limited applicability,
difficulty in describing complex chemistries and, inability to describe
systems where formation and breakage of bonds occur. Empirical force
fields are highly parameterized and typically include terms for bonds,
angles and torsions plus non-bonding terms {[}2{]}. Empirical force
fields are limited to the systems used in developing the parameters
(ex. proteins, lipids, DNA/RNA, etc) and the parameterizations usually
cover sp, sp\textsuperscript{2} and sp\textsuperscript{3} hybridizations.
It is extremely difficult to cover complex chemical spaces with force
fields, for example when transition metals are involved. The harmonic
nature of force fields does not typically allow for breaking and formation
of chemical bonds. In contrast, high-level quantum chemical methods
can describe most systems, but are still limited to small models,
at least when compared to typical systems studied by classical force
fields. QM/MM mix quantum chemical methods with empirical force fields
and, thus, are able to describe large systems. QM/MM methods work
better when the quantum region is highly localized but are inadequate
to describe the dynamics of large systems. 

Currently, many areas such as biophysics, biochemistry, materials
science, nanomedicine, etc. cannot be described using existing methodologies.
These systems have huge chemical spaces that are impossible to cover
using existing empirical force field methods and are too big for current
Quantum Mechanics (QM) techniques, even the best linear scaling methods.
There is a clear \textquotedblleft capabilities gap\textquotedblright{}
in existing computational methodologies that need to be urgently addressed.
Emerging fields such as nanomedicine or materials science would benefit
from new computational methodologies based on QM. Traditional applications
of classical force field methods would benefit as well. For example,
it is estimated that half of all proteins are metalloproteins {[}3{]}.
Simulations of metalloproteins would greatly benefit from fast QM
methods since existing classical force fields have problems describing
such systems.

In summary, new methodologies are needed to bridge the \textquotedblleft capabilities
gap\textquotedblright{} between current quantum chemical methods and
classical force fields. In the base Hartree-Fock method, the major
contributors to the cost of the calculation are the computation of
the two-Electron-Repulsion Integrals (ERIs), with a quartically scaling
O(N\textsuperscript{4}), diagonalization of the Fock matrix with
a cubically scaling O(N\textsuperscript{3}) and the self-consistent
procedure that typically adds more than 10 iterations. Development
of new computational methodologies based on QM will have to address
each of the restrictions in order to achieve acceptable speeds. The
aim of the current work is to develop an alternative technique, based
on accurate numerical approximations, for the fast computation of
ERIs. In computational quantum chemistry, the most common basis sets
are based on Gaussian basis functions. It was already apparent in
the 1950s that calculations of polyatomic systems based on Slater-type
orbitals would be intractable. The breakthrough occurred when Boys
proposed basis functions based on Cartesian Gaussian functions {[}4{]}.
It was also found that linear combinations of Gaussians, designated
as contracted Gaussians, could approximate atomic orbitals with great
accuracy. Ever since, contracted Gaussians have been the basis set
of choice, being used in all of the major program packages. 

Computation of ERIs has a long history. Initially, all molecular integrals
were calculated analytically since closed formulas for integrals over
Gaussians were easily derived. The analytical formulas being specific
to each integral do not allow the systematic calculation of integrals
of higher angular momentum. Several recursive methodologies were then
developed and gained acceptance in modern computational quantum chemistry
programs. In this category are included the methods of Rys polynomials
{[}5,6{]}, McMurchie and Davidson {[}7{]} and Obara and Saika {[}8{]}.
More recently, active work has been developed on approximate methodologies
to speed up the computation of ERIs, for example, approaches using
density fittings or the Cholesky decomposition. A very good and recent
review of the calculation of ERIs has been published by Reine \textit{et
al.} {[}9{]}. 

The methodology to compute ERIs proposed in this work differs in concept
and praxis relative to previous and current approaches. Existing methodologies
need to be generic and applicable to any basis set. In contrast, the
method being proposed approximates a pre-determined basis set and
is optimized for speed, as it needs to be many orders of magnitude
faster than current methods. The new computational methodology will
use single-\textgreek{z} (double-\textgreek{z} for transition metals)
basis sets. This paper is a proof-of-concept on the development of
accurate numerical approximations to the analytical formulae for ERIs.
The work will focus on the integral $\left(p_{xA}p_{xB}|p_{xC}p_{xC}\right)$
with atoms A, B, and C being carbon. The choice of three-center integrals
offers significant advantages. The numerical approximations are simpler
in three-center ERIs than in four-center ERIs because a smaller number
of coordinates are required. By keeping all elements the same, significant
symmetry relationships are introduced and smaller domains for the
coordinates can be considered (see Sect. \ref{sec:Results-and-discussion 4};
Figure S1), thus reducing the number of target points that are required
for the numerical approximations.

\section{\label{sec:Theoretical background 2}Theoretical background }

\subsection{\label{sub:A. Revisit the analytical calculation of two-electron-repulsion integrals 2A}Revisit
the analytical calculation of two-electron-repulsion integrals }

In the early years of computational chemistry, ERIs were calculated
using analytical formulae {[}10{]}. The notation used for the explicit
expressions of ERIs over Cartesian Gaussian functions is kept as close
as possible to the one used by Clementi {[}11{]}. An important concept
in molecular orbital theory is the expansion of the basis functions
$\phi_{i}\left(A\right)$ as liner-combinations of primitive Cartesian
Gaussian Type functions (GTFs):

\begin{equation}
\phi_{i}\left(A\right)={\displaystyle \stackrel[a=1]{n}{\sum}c_{ia}}\eta_{a}\left(A\right)
\end{equation}

Cartesian GTFs are composed of a radial Gaussian function multiplied
by Cartesian coordinates x,y and z with exponents l\textsubscript{i},
m\textsubscript{i}, and n\textsubscript{i}, 

\begin{equation}
\eta_{i}\left(A\right)=x_{A}^{l_{i}}y_{A}^{m_{i}}z_{A}^{n_{i}}\exp\left(-\alpha_{i}r_{A}^{2}\right)
\end{equation}

The basic steps required to derive the ERI of the kind $\left(p_{xA}p_{xB}|p_{xC}p_{xC}\right)$
are briefly described. Testing the novel numerical algorithms on three-center
ERIs is important because they are significantly simpler than the
four-center counterparts, due to having fewer degrees of spatial freedom,
while still requiring the same techniques to perform the approximation.
ERIs over basis functions are themselves written as linear combinations
over the primitive GTFs: 

\begin{equation}
\left(\phi_{A}\text{\ensuremath{\phi}}_{B}\text{|\ensuremath{\phi}}_{C}\phi_{C}\right)=\underset{a,b,c,c}{\sum}c_{a}c_{b}c_{c}c_{c}\left(\eta_{A}\eta_{B}|\eta_{C}\eta_{C}\right)
\end{equation}

The advantage of using GTFs stems from the Gaussian product theorem
is that the product of two GTFs is another GTF. In Eq. 3 the product
of the first pair centered at $\overrightarrow{A}$ and $\overrightarrow{B}$
results in the general formula:

\begin{widetext}

\begin{multline}
\eta\left(\alpha_{1},\overrightarrow{A},l_{1}\right)\eta\left(\alpha_{2},\overrightarrow{B},l_{2}\right)=\exp\left(-\frac{\alpha_{1}\alpha_{2}(\overline{AB}){}^{2}}{\gamma_{1}}\right)\text{\texttimes}\stackrel[i=0]{l_{1}+l_{2}}{\sum}f_{i}\left(l_{1},l_{2},\overline{PA}{}_{x},\overline{PB}{}_{x}\right)x_{P}^{i}\exp\left(-\gamma_{1}x_{P}^{2}\right)\text{\texttimes}\\
\stackrel[j=0]{m_{1}+m_{2}}{\sum}f_{j}\left(m_{1},m_{2},\overline{PA}{}_{y},\overline{PB}{}_{y}\right)y_{P}^{j}\exp\left(-\gamma_{1}y_{P}^{2}\right)\text{\texttimes}\stackrel[k=0]{n_{1}+n_{2}}{\sum}f_{k}\left(k_{1},k_{2},\overline{PA}{}_{z},\overline{PB}{}_{z}\right)z_{P}^{k}\exp\left(-\gamma_{1}z_{P}^{2}\right)\label{eq:}
\end{multline}

\end{widetext}

with 

\begin{equation}
\gamma_{1}=\alpha_{1}+\alpha_{2}
\end{equation}

and 

\begin{equation}
\overrightarrow{P}=\frac{\alpha_{1}\overrightarrow{A}+\alpha_{2}\overrightarrow{B}}{\gamma_{1}}
\end{equation}

Similar equations can be derived for the second pair with: 

\begin{equation}
\gamma_{2}=\alpha_{3}+\alpha_{4}
\end{equation}

and 

\begin{equation}
\overrightarrow{Q}=\frac{\alpha_{3}\overrightarrow{C}+\alpha_{4}\overrightarrow{D}}{\gamma_{2}}
\end{equation}

The functions $f_{i},\text{\dots},f_{k}$ appearing in Eq. 4 result
from the application of the binomial theorem to the products of Gaussian
functions. Their generic formula is: 

\begin{multline}
f_{i}\left(l_{1},l_{2},A,B\right)=\\
\stackrel[j=\max\left(0,i-l_{2}\right)]{\min\left(i,l_{1}\right)}{\sum}\frac{l_{1}!l_{2}!A^{l_{1}-j}B^{l_{1}-j}}{j!\left(l_{1}-j\right)!\left(l_{2}-i+j\right)!\left(l_{2}-i+j\right)!}
\end{multline}

Explicit values of the function $f_{i}\left(l_{1},l_{2},A,B\right)$
are given in Table \ref{tab:Possible values of the f function as a function}
up to l\textsubscript{1}+l\textsubscript{2}=4. Substituting the
pairs $\eta\left(\alpha_{1},\overrightarrow{A},l_{1}\right)\eta\left(\alpha_{2},\overrightarrow{B},l_{2}\right)$
and $\eta\left(\alpha_{3},\overrightarrow{C},l_{3}\right)\eta\left(\alpha_{4},\overrightarrow{D},l_{4}\right)$
into Eq. 3 results in the formal formula for the analytical calculation
of ERIs (Eq. 10) where the normalization factors are written as $N_{\alpha}$: 

\begin{widetext}

\begin{multline}
\left(\phi_{A}\text{\ensuremath{\phi}}_{B}\text{|\ensuremath{\phi}}_{C}\phi_{C}\right)=\underset{a,b,c,c}{\sum}c_{a}c_{b}c_{c}c_{c}\left(\eta_{A}\eta_{B}|\eta_{C}\eta_{C}\right)=\underset{a,b,c,c}{\sum}c_{a}c_{b}c_{c}c_{c}\exp\left(-\frac{\alpha_{1}\alpha_{2}(\overline{AB}){}^{2}}{\gamma_{1}}\right)\exp\left(-\frac{\alpha_{3}\alpha_{4}(\overline{CD}){}^{2}}{\gamma_{2}}\right)\text{\texttimes}\\
\stackrel[i=0]{l_{1}+l_{2}}{\sum}f_{i}\left(l_{1},l_{2},\overline{PA}{}_{x},\overline{PB}{}_{x}\right)\text{\texttimes}\stackrel[j=0]{m_{1}+m_{2}}{\sum}f_{j}\left(m_{1},m_{2},\overline{PA}{}_{y},\overline{PB}{}_{y}\right)\text{\texttimes}\stackrel[k=0]{n_{1}+n_{2}}{\sum}f_{k}\left(k_{1},k_{2},\overline{PA}{}_{z},\overline{PB}{}_{z}\right)\text{\texttimes}\\
\stackrel[i^{'}=0]{l_{3}+l_{4}}{\sum}f_{i^{'}}\left(l_{3},l_{4},\overline{QC}{}_{x},\overline{QD}{}_{x}\right)\text{\texttimes}\stackrel[j^{'}=0]{m_{1}+m_{2}}{\sum}f_{j^{'}}\left(m_{3},m_{3},\overline{QC}{}_{y},\overline{QD}{}_{y}\right)\text{\texttimes}\stackrel[k^{'}=0]{n_{3}+n_{4}}{\sum}f_{k^{'}}\left(k_{3},k_{4},\overline{QC}{}_{z},\overline{QD}{}_{z}\right)\\
\iint x_{P}^{i}y_{P}^{j}z_{P}^{k}x_{Q}^{i^{'}}y_{Q}^{j^{'}}z_{P}^{k^{'}}\frac{1}{r_{12}}\exp\left(-\gamma_{1}r_{P_{1}}^{2}-\gamma_{2}r_{Q_{2}}^{2}\right)dV_{1}dV_{2}
\end{multline}

A simplified notation, $\left\{ x_{P_{1}}^{i}y_{P_{1}}^{j}z_{P_{1}}^{k}|x_{Q_{2}}^{i^{'}}y_{Q_{2}}^{j^{'}}z_{Q_{2}}^{k^{'}}\right\} $,
is introduced for the integral $\iint x_{P}^{i}y_{P}^{j}z_{P}^{k}x_{Q}^{i^{'}}y_{Q}^{j^{'}}z_{P}^{k^{'}}\frac{1}{r_{12}}\exp\left(-\gamma_{1}r_{P_{1}}^{2}-\gamma_{2}r_{Q_{2}}^{2}\right)dV_{1}dV_{2}$
in the remaining of the text. 

\end{widetext}

Calculation of ERIs according to Eq. 10 requires repeated evaluations
of $\left\{ x_{P_{1}}^{i}y_{P_{1}}^{j}z_{P_{1}}^{k}|x_{Q_{2}}^{i^{'}}y_{Q_{2}}^{j^{'}}z_{Q_{2}}^{k^{'}}\right\} $,
\textquotedbl{}f\textquotedbl{} functions, normalization factors and
the two exponential functions, $\exp\left(-\frac{\alpha_{1}\alpha_{2}(\overline{AB}){}^{2}}{\gamma_{1}}\right)$
and $\exp\left(-\frac{\alpha_{3}\alpha_{4}(\overline{CD}){}^{2}}{\gamma_{2}}\right)$,
over multiple loops. There are loops over the contraction coefficients,
c\textsubscript{a}, c\textsubscript{b}, c\textsubscript{c} and
c\textsubscript{d}, and the indices i, j, etc. The indices i, j,
etc. determine the f functions and the integrals $\left\{ x_{P_{1}}^{i}y_{P_{1}}^{j}z_{P_{1}}^{k}|x_{Q_{2}}^{i^{'}}y_{Q_{2}}^{j^{'}}z_{Q_{2}}^{k^{'}}\right\} $.
The index i runs between 0 and $l_{1}+l_{2}$ and similarly for j,
k, i\textquoteright , \dots , which depend on $m_{1}+m_{2}$, $n_{1}+n_{2}$,
$l_{3}+l_{4}$, \dots . When the exponents i, j, \dots{} are zero,
the corresponding $x_{P_{1}}^{i}$, $y_{P_{1}}^{j}$, \dots{} terms
are indicated as \textquotedblleft 1\textquotedblright{} in $\left\{ x_{P_{1}}^{i}y_{P_{1}}^{j}z_{P_{1}}^{k}|x_{Q_{2}}^{i^{'}}y_{Q_{2}}^{j^{'}}z_{Q_{2}}^{k^{'}}\right\} $.

Although the complexity of the integrals $\left\{ x_{P_{1}}^{i}y_{P_{1}}^{j}z_{P_{1}}^{k}|x_{Q_{2}}^{i^{'}}y_{Q_{2}}^{j^{'}}z_{Q_{2}}^{k^{'}}\right\} $
increases with larger values of l\textsubscript{1}, l\textsubscript{2},
\dots , each is a well-defined function of $\overrightarrow{P}$ and
$\overrightarrow{Q}$ through the distances $\overline{PQ}$ and the
corresponding non-zero projections along the Cartesian axis $\overline{PQ}_{x}$
(also $\overline{PQ}_{y}$, $\overline{PQ}_{z}$, when the integral
involve y- and z- functions). Recalling the definitions of $\overrightarrow{P}$
and $\overrightarrow{Q}$ from Eqs 6 and 8, respectively, the integrals
$\left\{ x_{P_{1}}^{i}y_{P_{1}}^{j}z_{P_{1}}^{k}|x_{Q_{2}}^{i^{'}}y_{Q_{2}}^{j^{'}}z_{Q_{2}}^{k^{'}}\right\} $
are functions of the coordinates of A, B, C, and D (A,B, and C in
the three-center case).

The main purpose of this work is to illustrate how the three-center
integrals $\left(p_{xA}p_{xB}|p_{xC}p_{xC}\right)$, when carbon atoms
are placed at the centers A, B, and C, can be calculated accurately
through numerical approximation. In the following Sections, the specific
simplifications introduced by the considering a three-center ERI,
and the mathematical details of the numerical approximations are discussed.

\subsection{\label{sub:B. Numerical fitting of three-center two-electron-repulsion integrals 2B}Numerical
fitting of three-center two-electron-repulsion integrals \textmd{\normalsize{}$\mathbf{\left(p_{xA}p_{xB}|p_{xC}p_{xC}\right)}$} }

Calculation of three-center ERI $\left(p_{xA}p_{xB}|p_{xC}p_{xC}\right)$
involves significant simplifications resulting from many of the \textquotedbl{}f\textquotedbl{}
functions becoming null, according to Table \ref{tab:Possible values of the f function as a function}.
After the null terms are omitted, Eq. 10 can be rewritten as:

\begin{widetext}
\begin{multline}
\left(p_{xA}p_{xB}|p_{xC}p_{xC}\right)=\underset{a,b,c}{\sum}c_{a}c_{b}c_{c}^{2}N_{a}N_{b}N_{c}^{2}\exp\left(-\frac{\alpha_{1}\alpha_{2}(\overline{AB}){}^{2}}{\gamma_{1}}\right)\text{\texttimes}\\
\left[\overline{PA}{}_{x}\overline{PB}{}_{x}\left\{ 111|x_{Q_{2}}^{2}11\right\} +\left(\overline{PA}{}_{x}+\overline{PB}{}_{x}\right)\left\{ x_{P_{1}}11|x_{Q_{2}}^{2}11\right\} +\left\{ x_{P_{1}}^{2}11|x_{Q_{2}}^{2}11\right\} \right]
\end{multline}

The expressions of $\left\{ x_{P_{1}}^{2}11|x_{Q_{2}}^{2}11\right\} $,
$\left\{ x_{P_{1}}11|x_{Q_{2}}^{2}11\right\} $, and $\left\{ 111|x_{Q_{2}}^{2}11\right\} $
are respectively: 

\begin{subequations}

\begin{gather}
\left\{ x_{P_{1}}^{2}11|x_{Q_{2}}^{2}11\right\} =\frac{\pi^{5\text{\textfractionsolidus}2}}{2\beta\left(\gamma_{1}+\gamma_{2}\right){}^{7/2}}\text{\texttimes}\\
\left\{ 4\beta^{2}F_{4}\left(t\right)\overline{PQ}_{x}^{4}-12\beta F_{3}\left(t\right)\overline{PQ}_{x}^{3}+\left[2\left(\gamma_{1}+\gamma_{2}\right)\overline{PQ}_{x}^{2}+3\right]F_{2}\left(t\right)-\frac{\left(\gamma_{1}+\gamma_{2}\right)}{\beta}F_{1}\left(t\right)+\frac{\left(\gamma_{1}+\gamma_{2}\right)}{\beta}F_{0}\left(t\right)\right\} \nonumber \\
\left\{ x_{P_{1}}11|x_{Q_{2}}^{2}11\right\} =\frac{\pi^{5\text{\textfractionsolidus}2}}{\beta\left(\gamma_{1}+\gamma_{2}\right){}^{7/2}}\left\{ 2\gamma_{2}\beta F_{3}\left(t\right)\overline{PQ}_{x}^{3}-\left[2\gamma_{2}F_{2}\left(t\right)+\left(\gamma_{1}+\gamma_{2}\right)F_{1}\left(t\right)\right]\overline{PQ}_{x}\right\} \\
\left\{ 111|x_{Q_{2}}^{2}11\right\} =\frac{\pi^{5\text{\textfractionsolidus}2}}{\beta\left(\gamma_{1}+\gamma_{2}\right){}^{7/2}}\left\{ 2\gamma_{2}^{2}F_{2}\left(t\right)\overline{PQ}_{x}^{2}-\frac{\gamma_{2}\left(\gamma_{1}+\gamma_{2}\right)}{\gamma_{1}}F_{1}\left(t\right)+\frac{\left(\gamma_{1}+\gamma_{2}\right)^{2}}{\gamma_{1}}F_{0}\left(t\right)\right\} \ ,
\end{gather}

\end{subequations}

\end{widetext} where \textgreek{b} is defined as $\frac{\gamma_{1}\gamma_{2}}{\left(\gamma_{1}+\gamma_{2}\right)}$
and the terms $F_{n}\left(t\right)$ are the Boys function: 

\begin{equation}
F_{n}\left(t\right)=\int_{0}^{1}x^{2n}\exp\left(-tx^{2}\right)dx
\end{equation}
The evaluation of the Boys function had a recent renewed interest
and was the subject of recent publications {[}12,13{]}. A different
algorithm was developed for this work and will be discussed in a forthcoming
paper.

The integrals $\left\{ x_{P_{1}}^{i}y_{P_{1}}^{j}z_{P_{1}}^{k}|x_{Q_{2}}^{i^{'}}y_{Q_{2}}^{j^{'}}z_{Q_{2}}^{k^{'}}\right\} $
have important characteristics that can be explored to simplify the
numerical approximations. The factors \textgreek{g}\textsubscript{1},
\textgreek{g}\textsubscript{2} and \textgreek{b} depend on the orbital
exponents and are unaffected by geometrical changes. The Boys functions
$F_{n}\left(t\right)$ depend on the orbital exponents and the separation
between points $\overrightarrow{P}$ and $\overrightarrow{Q}$ , being
independent of the spatial orientation of the system. The factor $\overline{PQ}_{x}$
(also y and z), which is the x component of the vector $\overline{PQ}$,
depends on the orientation of the system. The \textquotedbl{}f\textquotedbl{}
functions also introduce terms that depend on the orientation of the
system, $\overline{PA}{}_{x}\overline{PB}{}_{x}$ and $\left(\overline{PA}{}_{x}+\overline{PB}{}_{x}\right)$
(see Table \ref{tab:Possible values of the f function as a function}
and Eq. 11). The algorithm developed for the calculation of ERIs is
based on the multivariate numerical approximation of all functions
contributing to the integrals on the desired interval. The terms contributing
to Eq. 11 consisting of products of Eqs 12a-12c and their respective
\textquotedblleft f\textquotedblright{} terms from Table \ref{tab:Possible values of the f function as a function}
have complex spatial dependencies resulting from the Boys functions,
$\overline{PA}{}_{x}\overline{PB}{}_{x}$, $\left(\overline{PA}{}_{x}+\overline{PB}{}_{x}\right)$
and $\overline{PQ}_{x}$ terms. The strategy used in this work consists
in recasting the parcels making the total ERI in terms of simpler
functions which are products of the rotationally dependent functions
$\overline{PQ}_{x}$, $\overline{PA}{}_{x}\overline{PB}{}_{x}$ and
$\left(\overline{PA}{}_{x}+\overline{PB}{}_{x}\right)$, designated
as $g_{n}^{rot}$, and a rotationally invariant term, $G_{n}$. The
index n is the exponent of $\overline{PQ}_{x}$. In addition to the
$g_{n}^{rot}/G_{n}$ terms, there is an additional rotationally dependent
term derived from $\overline{PA}{}_{x}\overline{PB}{}_{x}$ , $g_{\overline{PA}{}_{x}\overline{PB}{}_{x}}^{rot}$.
The corresponding rotationally invariant term is desinated as $G_{n\overline{PA}{}_{x}\overline{PB}{}_{x}}$.
Using the terms of $\overline{PQ}_{x}^{4}$ for illustration, the
rotationally dependent functions $g_{4}^{rot}$ and the corresponding
rotationally independent term $G_{4}$ are calculated as:

\begin{widetext}

\begin{equation}
g_{4}^{rot}\left(\overline{PQ}_{x}^{4}\right)=\frac{\underset{a,b,c}{\sum}c_{a}c_{b}c_{c}^{2}N_{a}N_{b}N_{c}^{2}\exp\left(-\frac{\alpha_{1}\alpha_{2}(\overline{AB}){}^{2}}{\gamma_{1}}\right)\times\frac{4\pi^{5\text{\textfractionsolidus}2}}{\beta\left(\gamma_{1}+\gamma_{2}\right){}^{7/2}}\times\left(\overline{PQ}_{x}^{4}\right)\times F_{4}\left(t\right)}{\underset{a,b,c}{\sum}c_{a}c_{b}c_{c}^{2}N_{a}N_{b}N_{c}^{2}\exp\left(-\frac{\alpha_{1}\alpha_{2}(\overline{AB}){}^{2}}{\gamma_{1}}\right)\text{\texttimes}\frac{4\pi^{5\text{\textfractionsolidus}2}}{\beta\left(\gamma_{1}+\gamma_{2}\right){}^{7/2}}\text{\texttimes}F_{4}\left(t\right)}
\end{equation}

\begin{equation}
G_{4}=\underset{a,b,c}{\sum}c_{a}c_{b}c_{c}^{2}N_{a}N_{b}N_{c}^{2}\exp\left(-\frac{\alpha_{1}\alpha_{2}(\overline{AB}){}^{2}}{\gamma_{1}}\right)\text{\texttimes}\frac{4\pi^{5\text{\textfractionsolidus}2}}{\beta\left(\gamma_{1}+\gamma_{2}\right){}^{7/2}}\text{\texttimes}F_{4}\left(t\right)
\end{equation}

The important rotationally invariant term $G_{0}$, which makes a
direct contribution to the total computed ERI, is 

\begin{multline}
G_{0}=\frac{\pi^{5\text{\textfractionsolidus}2}}{\beta\left(\gamma_{1}+\gamma_{2}\right){}^{7/2}}\text{\texttimes}\\
\left[6F_{2}(t)-\frac{2\left(\gamma_{1}+\gamma_{2}\right)}{\beta}F_{1}\left(t\right)+\frac{2\left(\gamma_{1}+\gamma_{2}\right)}{\beta}F_{0}\left(t\right)-\frac{\gamma_{2}\left(\gamma_{1}+\gamma_{2}\right)}{\gamma_{1}}F_{1}\left(t\right)+\frac{\left(\gamma_{1}+\gamma_{2}\right)^{2}}{\gamma_{1}}F_{0}\left(t\right)\right]
\end{multline}

\end{widetext}

\begin{table}
\caption{\label{tab:Possible values of the f function as a function}Possible
values of the f function as a function of l\protect\textsubscript{1},
l\protect\textsubscript{2} and the generic parameters A and B.}

\begin{tabular}{|c|c|c|}
\hline 
\multicolumn{3}{|c|}{Eq. 9}\tabularnewline
\hline 
\hline 
 & $A\neq0,B\neq0$ & $A=0,B=0$\tabularnewline
\hline 
\multicolumn{3}{|c|}{$l_{1}=2,l_{2}=2$}\tabularnewline
\hline 
$i=0$ & $A^{2}B^{2}$ & $0$\tabularnewline
\hline 
$i=1$ & $2AB^{2}+2A^{2}B$ & $0$\tabularnewline
\hline 
$i=2$ & $B^{2}+4AB+A^{2}$ & $0$\tabularnewline
\hline 
$i=3$ & $2B+2A$ & $0$\tabularnewline
\hline 
$i=4$ & $1$ & $1$\tabularnewline
\hline 
\multicolumn{3}{|c|}{$l_{1}=2,l_{2}=1$}\tabularnewline
\hline 
$i=0$ & $A^{2}B$ & $0$\tabularnewline
\hline 
$i=1$ & $2AB+A^{2}$ & $0$\tabularnewline
\hline 
$i=2$ & $B+2A$ & $0$\tabularnewline
\hline 
$i=3$ & $1$ & $1$\tabularnewline
\hline 
\multicolumn{3}{|c|}{$l_{1}=2,l_{2}=0$}\tabularnewline
\hline 
$i=0$ & $A^{2}$ & $0$\tabularnewline
\hline 
$i=1$ & $2A$ & $0$\tabularnewline
\hline 
$i=2$ & $1$ & $1$\tabularnewline
\hline 
\multicolumn{3}{|c|}{$l_{1}=1,l_{2}=1$}\tabularnewline
\hline 
$i=0$ & $AB$ & $0$\tabularnewline
\hline 
$i=1$ & $B+A$ & $0$\tabularnewline
\hline 
$i=2$ & $1$ & $1$\tabularnewline
\hline 
\multicolumn{3}{|c|}{$l_{1}=1,l_{2}=0$}\tabularnewline
\hline 
$i=0$ & $A$ & $0$\tabularnewline
\hline 
$i=1$ & $1$ & $1$\tabularnewline
\hline 
\multicolumn{3}{|c|}{$l_{1}=0,l_{2}=0$}\tabularnewline
\hline 
$i=0$ & $1$ & $1$\tabularnewline
\hline 
 &  & \tabularnewline
\hline 
\end{tabular}
\end{table}

\section{\label{sec:Mathematical background 3}Mathematical background }

\subsection{\label{sub:Multivariate-approximation 3A}Multivariate approximation }

In many applications, it is convenient to introduce \textit{approximate
functions}. An approximate function $g\left(x\right)$ is a function
that given $m$ data points $x$ approximates the target values produced
by the function $f\left(x\right)$ as closely as possible according
to some metric. The approximant $g\left(x\right)$ is desired to be
as smooth and compact as possible. The need to approximate often occurs
when it is too costly or complex to use the true function, or even
when the true function is unknown. The mathematical theory of approximation
is well documented (see for example Ref. {[}14{]}). This work explores
the possibility of approximating the complex and computationally costly
Eq. 11 with simpler, and faster to evaluate, functions. All approximants
are based on polynomial expansions (Eq. 17), in which the coefficients
$a_{i}$ are scalars and the generic basis functions $H\left(x\right)$
can take different forms: 

\begin{equation}
f\left(x\right)=a_{0}+a_{1}H_{1}\left(x\right)+\cdots+a_{n}H_{n}\left(x\right)
\end{equation}
The main criterion to determine the quality of an approximation is
the measurement of the \textquotedblleft distance\textquotedblright{}
between the target data points and the same set of points as obtained
by the specified approximating function (approximant). It is important
that the target and approximated points are as close as possible.
A suitable metric to account for the global different between the
set of true values and their respective approximations used extensively
in this work is the l\textsubscript{2}-norm. 

The multivariate approximation scheme developed to approximate ERIs
consists of multiple levels of bivariate (or univariate) approximants,
with the fitting variables of a given level being expressed in terms
of the variables of the next immediate level. The methodology is illustrated
with the help of a 3-dimensional model depicted in Figure 1. To approximate
the point$f(x_{1},x_{2},x_{3})$, represented by the red sphere, a
numerical approximant $g(x_{1},x_{2})$ of all points on the $(x_{1},x_{2})$
plane is first developed. The function $g\left(x\right)$ is expanded
in terms of primitive functions $H_{n}\left(x_{1},x_{2}\right)$ according
with Eq. 17. The dependency of $x_{3}$, which is illustrated in Figure
1 by the vector originating at the blue sphere, is carried by fitting
parameters $a_{i}$ as functions of $x_{3}$. In mathematical terms,
the dependency of the fitting parameters $a_{i}$ is given by another
expansion similar to Eq. 17. The basis functions are represented by
$H_{n}^{'}(x_{3})$ and the expansion has adjustable coefficients
$a_{i}$: 

\begin{equation}
a\left(x_{3}\right)=a_{0}^{'}+\stackrel[i=1]{n'}{\sum}a_{i}^{'}H_{i}^{'}\left(x_{3}\right)
\end{equation}
The process can be repeated multiple times, generating complex dependencies
of multivariate functions. However, as the number of fitting parameters
grow very fast with each additional layer of variable dependencies,
in practice, the process is limited to a small number of layers.

\begin{figure}
\includegraphics[scale=0.7]{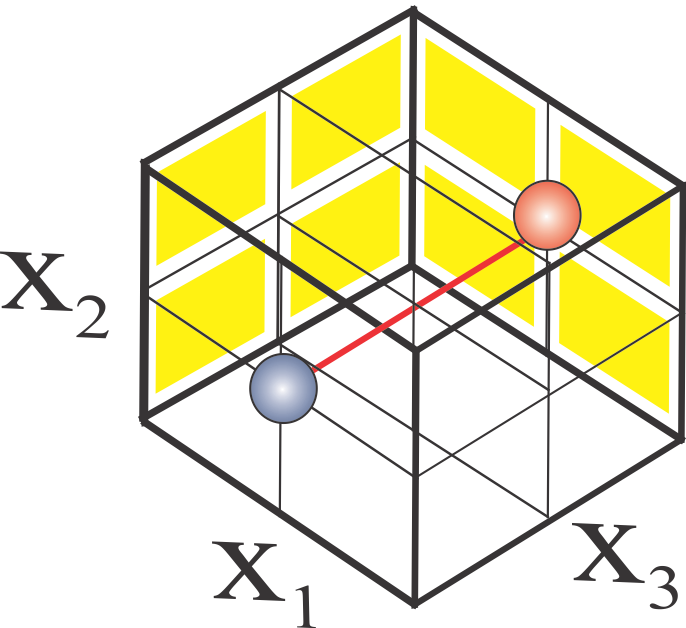}\caption{Illustration of the spatial dependency of multilayered approximating
functions.}
\end{figure}

In this work, the need for smooth functions arises because of the
multiple dependencies of the variables. Since the fitting parameters
carry additional dependencies themselves it is important that they
are as smooth as possible to avoid discontinuities that make the next
level fittings more complex. Other important criteria in defining
the fitting process are computational efficiency, simplicity of algorithm
implementation and future evolution of the method. When designing
algorithms for numerical approximation is important to consider how
fast and accurate the method is in the present, and to have a clear
plan for future development.

The fitting functions were chosen to be bivariate Chebyshev orthogonal
polynomials. Chebyshev polynomials form an important class of functions
in curve fitting {[}15{]}. A similar expansion can be developed for
surfaces $f\left(x,y\right)$ where the polynomial is based on to
Chebyshev series with $\overline{x},\overline{y}\in[-1,1]\times[-1,1]$:

\begin{equation}
f(x,y)=\stackrel[i=0]{n_{x}}{\sum}\stackrel[j=0]{n_{y}}{\sum}a_{ij}T_{i}\left(\overline{x}\right)T_{j}\left(\overline{y}\right)
\end{equation}
The arguments $\overline{x}$ and $\overline{y}$ are obtained from
the original variables x and y by the transformations 

\begin{equation}
\overline{x}=\frac{2x-\left(x_{max}+x_{min}\right)}{x_{max}-x_{min}}
\end{equation}
and 

\begin{equation}
\overline{y}=\frac{2y-\left(y_{max}+y_{min}\right)}{y_{max}-y_{min}}
\end{equation}
The two-dimensional Chebyshev expansion was evaluated directly by
computing the polynomials and summing all contributions according
to Eq. 19.

\subsection{\label{sub:Choice-of-coordinates 3B}Choice of coordinates }

Each of the terms $g_{n}^{rot}$ and $G_{n}$ required to calculate
ERIs according to the prescription of Eqs 14-16 can be expressed in
terms of a finite number of variables. Fitting of three-center ERIs
requires six coordinates that are used to position the atomic centers
carrying the basis functions. Importantly, only the total number of
variables has to be fulfilled and not the nature of the individual
coordinates as long as they provide the spatial assignment of the
atomic centers. It is, however, advisable to use combinations of variables
that lead to simpler fitting expressions, in addition to having physical
meanings that can be related to common geometrical transformations.
In this respect bond distances, angles and torsions are prime candidates. 

\begin{figure}
\includegraphics[scale=0.3]{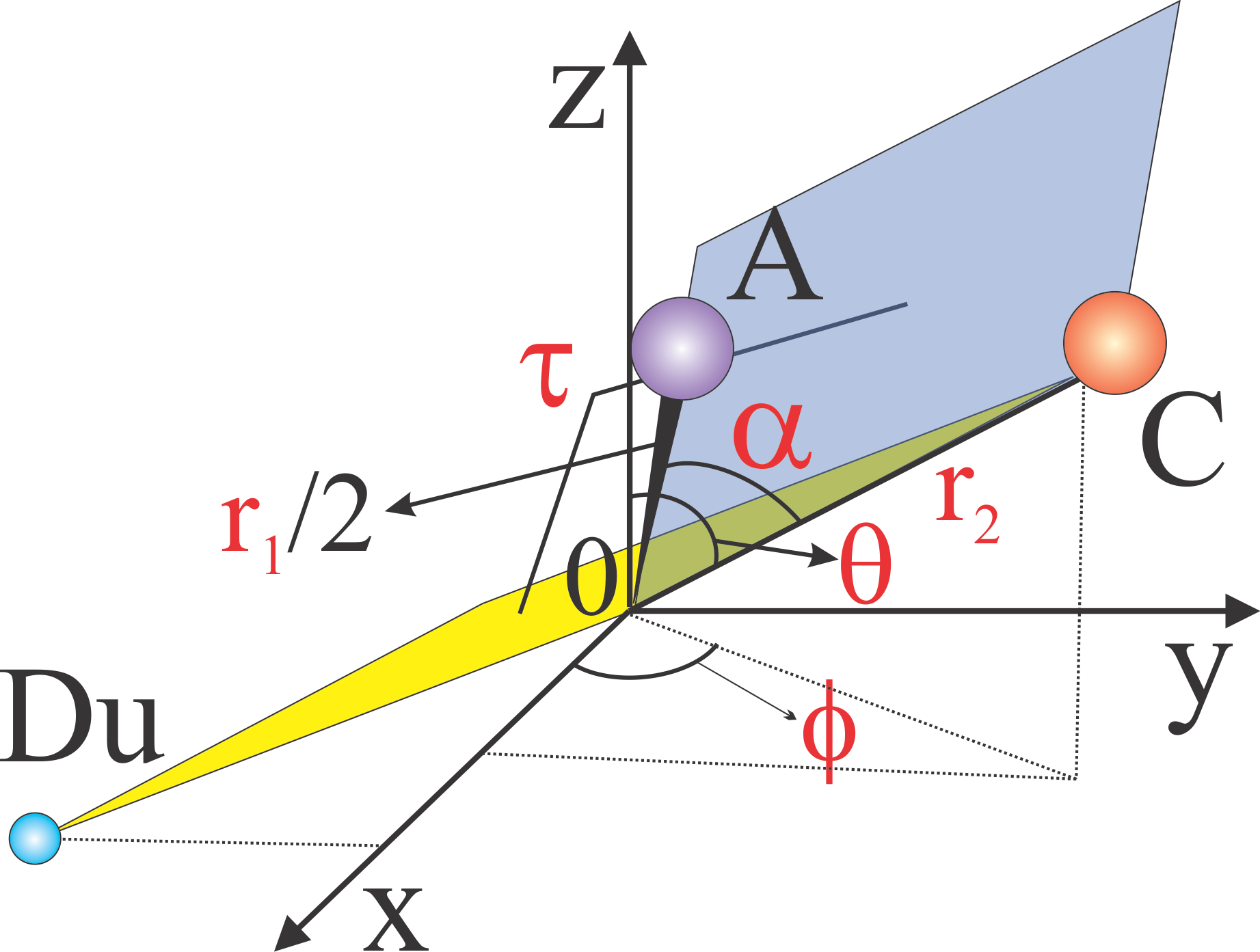}\caption{Illustration of the coordinates used in the fitting of three-center
electron repulsion integrals}
\end{figure}

The protocol followed in this work consists in separating the rotationally
dependent terms of $\overline{PQ}_{x}^{n}$ and $\overline{PA}_{x}\overline{PB}_{x}$
from the rotational invariant counterparts. The set of coordinates
chosen for the fitting of the rotationally invariant part are two
distances, r\textsubscript{1} and r\textsubscript{2}, and the internal
angle \textgreek{a}. r\textsubscript{1} is the distance between the
centers A and B and r\textsubscript{2} is the separation between
C and the midpoint of $\overrightarrow{AB}$ represented by $\overrightarrow{O}$.
\textgreek{a} is the angle $\hat{COA}$ (see Figure 2). The projections
$\overline{PQ}_{x}$ , $\overline{PQ}_{y}$ or $\overline{PQ}_{z}$
require special attention since they impart the rotational invariance
of the integrals. Their spatial dependencies are significantly more
complex, requiring an extra set of coordinates. The extra variables
are the polar spherical coordinates $\theta$ and $\varphi$, which
are used to position the atomic center C and the dihedral angle $\tau$
,which is used to determine the relative orientation of centers A
(and B) relative to C (see Figure 2).

\section{\label{sec:Results-and-discussion 4}Results and discussion }

The following Section is dedicated to evaluating the accuracy and
speed of the numerical algorithm to approximate ERIs. Emphasis is
placed on testing the ability of the method to accurately reproduce
the integral $\left(p_{xA}p_{xB}|p_{xC}p_{xC}\right)$ with carbon
atoms are placed at A, B, and C positions. All calculations were based
on the STO-6G basis set. This basis set is sufficiently small to allow
computation of the many target ERIs used in the parameterization in
a reasonable time. All calculations were done on modest hardware:
AMD 8350 CPU and 24 GByte of RAM memory. No parallelization was attempted
and the calculations were done on a single-core. All codes were compiled
with Gfortran using the \textendash O3 compiler flag. The approximation
of the rotationally independent terms is discussed first, with $G_{4}$
being used as example. Afterward, the fitting of the rotationally
dependent terms is analyzed. The approximating methodologies are illustrated
with the help of $g_{4}^{rot}\left(\overline{PQ}_{x}^{4}\right)$
since it is representative of the other terms. The accuracy and speed
of the multivariate methodology of approximation are discussed in
Sect. \ref{sub:B. Fitting the rotationally dependent terms 4B} and
\ref{sub:Adding-all-together 4C}. Two quantities are used to measure
the goodness-of-fit of the approximants: the Root Mean Square Error
(RMSE) and R\textsuperscript{2}. RMSE measures the total deviation
of the computed from the target values, and a value closer to zero
indicates the fit is good and is useful for prediction. Another quantity
to access the quality of an approximation is R\textsuperscript{2},
which indicates how well the approximation explains variation in the
data. The closest the value of R\textsuperscript{2} is to one the
better is the approximation. The domains of the variables influencing
the rotation of the systems are $\alpha\in\left[0,180\text{\textdegree}\right]$,
and $\tau,\theta,\varphi\text{\ensuremath{\in}}\left[0,90\text{\textdegree}\right]$.
The domains of $\tau$, $\theta$, and $\varphi$ are limited to 90º
because of the symmetry relations resulting from having the same element
at the positions A, B, and C. Figure S1 illustrates the symmetry effects
for the dependencies of $\left(\alpha,\tau\right)$ and $\left(\theta,\varphi\right)$
for the function $g_{4}^{rot}\left(\overline{PQ}_{x}^{4}\right)$.

\subsection{\label{sub:A. Fitting the rotationally independent terms 4A}Fitting
the rotationally independent terms $\mathbf{G_{n}\left(\alpha,r_{1},r_{2}\right)}$
and $\mathbf{G_{\overline{PA}_{x}\overline{PB}_{x}}\left(\alpha,r_{1},r_{2}\right)}$}

The protocol described in Sect. \ref{sub:Multivariate-approximation 3A}
for the multivariate fitting of the different parcels making the analytical
expression of the ERIs starts with the initial fitting of rotationally
invariant functions $G_{n}\left(\alpha,r_{1},r_{2}\right)$ and $G_{\overline{PA}_{x}\overline{PB}_{x}}\left(\alpha,r_{1},r_{2}\right)$.
In most cases, these are auxiliary functions used to create smoother
rotationally dependent surfaces that are easier to fit, although $G_{0}\left(\alpha,r_{1},r_{2}\right)$
contributes directly to the final integral (Eq. 16). The dependencies
of $G_{n}$ are on the angle \textgreek{a} and distances r\textsubscript{1}
and r\textsubscript{2}. The protocol for the fitting of $G_{n}\left(\alpha,r_{1},r_{2}\right)$
and $G_{\overline{PA}_{x}\overline{PB}_{x}}\left(\alpha,r_{1},r_{2}\right)$calls
to the initial fitting of the \textgreek{a} dependency. The plots
of the $G_{n}\left(\alpha,r_{1},r_{2}\right)$ functions relative
to \textgreek{a} (with r\textsubscript{1} and r\textsubscript{2}
fixed) follows a similar symmetric sinusoidal curve. The dependency
of $G_{n}\left(\alpha,r_{1}=2,6a.u.,r_{2}=5.0a.u.\right)$ is illustrated
on Figure S2. The function of choice for the fitting of the dependency
of \textgreek{a}, in radians, was a 10\textsuperscript{th}-order
polynomial written in the form: 

\begin{multline}
f\left(\alpha\right)=\\
a_{0}\left(r_{1},r_{2}\right)+a_{1}\left(r_{1},r_{2}\right)\alpha+\cdots+a_{10}\left(r_{1},r_{2}\right)\alpha^{10}
\end{multline}
The dependency of each $a_{i}$ term on $\left(r_{1},r_{2}\right)$
is highlighted in Eq. 22, in accordance with the multivariate fitting
algorithm described in Sect. \ref{sub:Multivariate-approximation 3A}.
The 10\textsuperscript{th} order expansion was found to be acceptable
in terms of computational cost and accuracy. Because of the dependence
of the coefficients on the distances r\textsubscript{1} and r\textsubscript{2},
it is important to keep the polynomial expansion above as compact
as possible.

\begin{figure}
\includegraphics[scale=0.45]{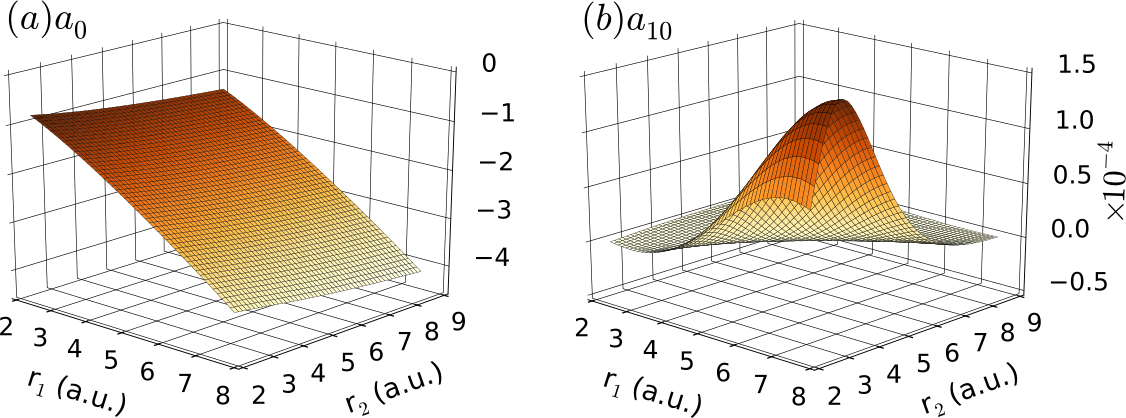}

\caption{Illustration of the spatial dependency of a\protect\textsubscript{0},
and a\protect\textsubscript{10} of the rotationally independent term
$G_{4}\left(\alpha,r_{1},r_{2}\right)$. The surfaces are smooth and
suitable for accurate approximation. The units of the surfaces are
radians\protect\textsuperscript{-1}.}
\end{figure}

The dependency of the coefficients a\textsubscript{0} and a\textsubscript{10}
in Eq. 22 on $\left(r_{1},r_{2}\right)$ is illustrated in Figure
3. The key assumption of this work, that the fitting coefficients
of a polynomial approximant at a certain level have smooth spatial
dependencies of the variables of the next level, and thus, are able
to carry that spatial dependency, is fully fulfilled. Although no
rigorous mathematical proof is presented, the surfaces of a\textsubscript{0},
and a\textsubscript{10}, as well as the surfaces of the remaining
coefficients, are smooth and can be approximated using the multivariate
techniques presented before. Each coefficient $a_{i}\left(r_{1},r_{2}\right)$
was fitted by double Chebyshev polynomials with arguments $\overline{r}_{1}$
and $\overline{r}_{2}$ (see Eq. 19). The order of the expansion was
truncated at 14. It is the largest Chebyshev polynomial order used
in this work. The computational cost of using such a long expansion
is not prohibitive for two reasons. First, each increase of the Chebyshev
polynomial order only contributes the additional number of parameters
times the order of the univariate polynomial expansion of \textgreek{a}.
Second, the rotationally invariant terms only need to be calculated
once and stored. The same rotationally invariant terms, $G_{n}\left(\alpha,r_{1},r_{2}\right)$
and $G_{\overline{PA}_{x}\overline{PB}_{x}}\left(\alpha,r_{1},r_{2}\right)$,
can be used in the fitting of all ERIs whether they involve p\textsubscript{x},
p\textsubscript{y} or p\textsubscript{z} functions. The explicit
expression for fitting all spatial dependencies of $G_{n}\left(\alpha,r_{1},r_{2}\right)$,
combining Eq. 22 above with the bivariate $\left(r_{1},r_{2}\right)$
Chebyshev polynomial for each of the coefficients $a_{i}$ is 

\begin{widetext}

\begin{multline}
G_{n}\left(\alpha,r_{1},r_{2}\right)/G_{\overline{PA}_{x}\overline{PB}_{x}}\left(\alpha,r_{1},r_{2}\right)=\\
\left(\stackrel[i=0]{n_{x}}{\sum}\stackrel[j=0]{n_{y}}{\sum}a_{ij}^{(0)}T_{i}\left(\overline{r}_{1}\right)T_{j}\left(\overline{r}_{2}\right)\right)+\left(\stackrel[i=0]{n_{x}}{\sum}\stackrel[j=0]{n_{y}}{\sum}a_{ij}^{(1)}T_{i}\left(\overline{r}_{1}\right)T_{j}\left(\overline{r}_{2}\right)\right)\alpha+\cdots+\\
\left(\stackrel[i=0]{n_{x}}{\sum}\stackrel[j=0]{n_{y}}{\sum}a_{ij}^{(9)}T_{i}\left(\overline{r}_{1}\right)T_{j}\left(\overline{r}_{2}\right)\right)\alpha^{9}+\left(\stackrel[i=0]{n_{x}}{\sum}\stackrel[j=0]{n_{y}}{\sum}a_{ij}^{(10)}T_{i}\left(\overline{r}_{1}\right)T_{j}\left(\overline{r}_{2}\right)\right)\alpha^{10}
\end{multline}

\end{widetext} where $T_{i}\left(\overline{r}_{1}\right)$ is the
Chebyshev polynomial of the first kind of degree i with argument $\overline{r}_{1}$,
and $T_{j}\left(\overline{r}_{2}\right)$ is similarly defined for
j and $\overline{r}_{2}$. The approximations are extremely accurate
with overall RMSEs lower than 7.35E-05 and R\textsuperscript{2} coefficients
>\textcompwordmark{}> 0.99999. The residuals for $G_{n}\left(\alpha,r_{1},r_{2}\right)$
and $G_{\overline{PA}_{x}\overline{PB}_{x}}\left(\alpha,r_{1},r_{2}\right)$
are plotted in Figure S3 for $\alpha=145^{o}$.

\subsection{\label{sub:B. Fitting the rotationally dependent terms 4B}Fitting
the rotationally dependent terms $\mathbf{g_{n}^{rot}\left(\overline{PQ}_{x}^{n}\right)}$
and $\mathbf{g_{\overline{PA}{}_{x}\overline{PB}{}_{x}}^{rot}}$ }

The rotationally dependent functions hold the effects of the $\overline{PQ}_{x}^{n}$,
$\overline{PA}{}_{x}\overline{PB}{}_{x}$ and $\left(\overline{PA}{}_{x}+\overline{PB}{}_{x}\right)$
terms in the three-center ERI of the kind $\left(p_{xA}p_{xB}|p_{xC}p_{xC}\right)$.
These are considerably more challenging to approximate since they
depend on six variables instead of the three variables in the rotationally
independent terms. Two quantities were defined to measure the contribution
of each term to the total ERI: the Average Absolute Percentage Contribution
(AAPC) and Maximum Absolute Percentage Contribution (MAPC). The absolute
value of each term was chosen because each can be positive or negative.
The AAPC and MAPC quantities are calculated for a generic term f\textsubscript{a}
as respectively $100\times\stackrel[i]{n}{\sum}\frac{f_{a,i}}{\left(f_{1,i}+\cdots+f_{k,i}\right)n}$
and $100\times\max\frac{f_{a,i}}{\left(f_{1,i}+\cdots+f_{k,i}\right)}$.
Values closer to 100\% indicate a stronger contribution to the integral
and likewise values close to zero mean smaller contributions. The
most significant contributors are the rotationally independent term
$G_{0}$ and the rotationally dependent function of $\overline{PA}{}_{x}\overline{PB}{}_{x}$.
Interestingly, all the terms containing the projections $\overline{PQ}_{x}$
make considerably smaller contributions (see Table \ref{tab:Average Absolute Percentage Contribution (AAPC)}).
According to the relevance of each term, different expansions can
be defined without imparting significantly the accuracy of the final
approximation. The $G_{0}$ term is already fitted with the highest
order of any Chebyshev polynomial, and the accuracy of the approximations
can be hardly improved. 

\begin{table*}
\caption{\label{tab:Average Absolute Percentage Contribution (AAPC)}Average
Absolute Percentage Contribution (AAPC) and Maximum Absolute Percentage
Contribution (AAPC) of each term contributing to $\left(p_{xA}p_{xB}|p_{xC}p_{xC}\right)$}

\begin{tabular}{|c|c|c|c|c|c|c|c|}
\hline 
 & \multicolumn{7}{c|}{Term}\tabularnewline
\hline 
\hline 
 & $\overline{PQ}_{x}^{4}$ & $\overline{PQ}_{x}^{3}$ & $\overline{PQ}_{x}^{2}\left(l_{1}+l_{2}=0\right)$ & $\overline{PQ}_{x}$ & $\overline{PQ}_{x}^{2}\left(l_{1}+l_{2}=2\right)$ & $\overline{PA}{}_{x}\overline{PB}{}_{x}$ & $G_{0}$\tabularnewline
\hline 
AAPC & 0.4  & 0.1 & 0.8 & 0.4 & 1.0 & 45.7 & 51.6\tabularnewline
\hline 
MAPC  & 12.7 & 2.2 & 10.1 & 2.5 & 11.0 & 91.2 & 100.0\tabularnewline
\hline 
\end{tabular}
\end{table*}

The fitting protocol for $g_{n}^{rot}\left(\overline{PQ}_{x}^{n}\right)$
and \textbf{$g_{\overline{PA}{}_{x}\overline{PB}{}_{x}}^{rot}$} requires
three layers of fittings using bivariate Chebyshev polynomials. The
pairing of variables is: $\left(\theta,\varphi\right)$ $\rightarrow$
$\left(\alpha,\tau\right)$ $\rightarrow$ $\left(r_{1},r_{2}\right)$.
In the first step, fitting functions $f\left(\theta,\varphi\right)$
are determined for each of the target points $\left(\alpha,\tau,r_{1},r_{2}\right)$
(see Eq. 24a). The coefficients $a_{ij}$ carry the dependency of
the remaining variables $\alpha$, $\tau$, r\textsubscript{1}, and
r\textsubscript{2}. In the second level of optimization, each of
the coefficients $a_{ij}$ is fitted similarly with bivariate Chebyshev
polynomial (Eq. 24b). The fitting coefficients $b_{kl}^{ij}$ carry
the dependency of $\left(r_{1},r_{2}\right)$ and each will be fitted
in the third level of fittings (Eq. 24c).

\begin{widetext}

\begin{subequations}

\begin{gather}
g^{rot}\simeq f\left(\theta,\varphi\right)_{|\alpha^{0},\tau^{0},r_{1}^{0},r_{2}^{0}}=\stackrel[i=0]{n_{\theta}}{\sum}\stackrel[j=0]{n_{\varphi}}{\sum}a_{ij}\left(\alpha,\tau,r_{1},r_{2}\right)T_{i}\left(\overline{\theta}\right)T_{j}\left(\overline{\varphi}\right)
\end{gather}

\begin{gather}
a_{ij}\left(\alpha,\tau\right){}_{|r_{1}^{0},r_{2}^{0}}=\stackrel[k=0]{n_{\alpha}}{\sum}\stackrel[l=0]{n_{\tau}}{\sum}b_{kl}^{ij}\left(r_{1},r_{2}\right)T_{i}\left(\overline{\alpha}\right)T_{j}\left(\overline{\tau}\right)
\end{gather}

\begin{gather}
b_{kl}^{ij}\left(r_{1},r_{2}\right)=\stackrel[m=0]{n_{r_{1}}}{\sum}\stackrel[n=0]{n_{r_{2}}}{\sum}c_{mn}^{ij,kl}T_{i}\left(\overline{r}_{1}\right)T_{j}\left(\overline{r}_{2}\right)
\end{gather}

\end{subequations} 

In Eq. 24 the superscript \textquotedblleft 0\textquotedblright{}
means that the corresponding variable assumes a fixed value.

\end{widetext}

\begin{figure*}
\includegraphics[scale=0.6]{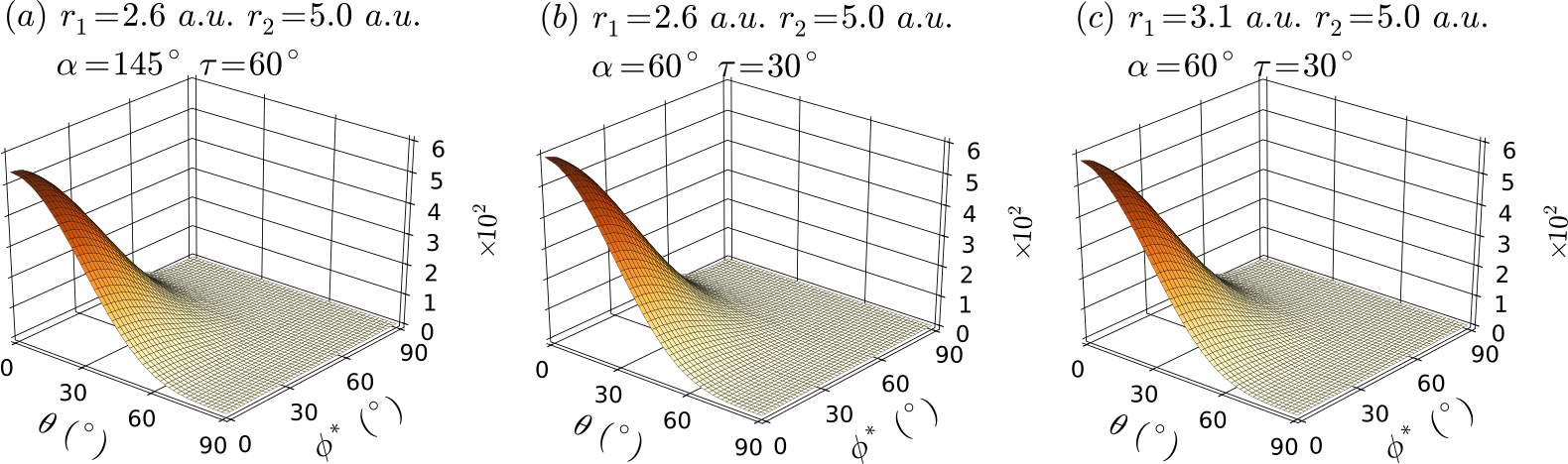}

\caption{Illustration of the dependency of $\theta$ and $\varphi$ for fixed
values of r\protect\textsubscript{1}, r\protect\textsubscript{2},
\textgreek{a}, and \textgreek{t} for $g_{4}^{rot}\left(\overline{PQ}_{x}^{4}\right)$.
(a) and (b) show the effect of varying \textgreek{a} and \textgreek{t},
whereas (b) and (c) illustrate the effect of varying r\protect\textsubscript{1},
r\protect\textsubscript{2}. Approximation of each surface will require
calculation of the coefficients a\protect\textsubscript{ij} of Equation
24a.}
\end{figure*}

Figure 4 illustrates selected surfaces $f\left(\theta,\varphi\right)$
for specific values of $\left(r_{1},r_{2}\right)$ and $\left(\alpha,\tau\right)$.
All surfaces have similar Gaussian-like shapes with maxima at $\left(0^{o},0^{o}\right)$.
It is noteworthy that to facilitate the numerical approximations,
the surfaces were symmetrized through the change of coordinate $\varphi*=180^{o}-\varphi$.
The functions of $\overline{PQ}_{x}^{n}$ were approximated with Chebyshev
polynomials of order 10. The rotationally dependent function of $\overline{PA}{}_{x}\overline{PB}{}_{x}$
was approximated with Chebyshev polynomials of order 10 and 11 and
the overall results are discussed in Sect. \ref{sub:Adding-all-together 4C}.

Despite the apparent similarity of the $f\left(\theta,\varphi\right)$
functions for fixed values of $\left(r_{1},r_{2}\right)$ the coefficients
$a_{ij}$ show remarkable variability as a function of \textgreek{a}
and \textgreek{t}. Figure 5 illustrates the coefficients a\textsubscript{1}
and a\textsubscript{66} for r\textsubscript{1} = 2.6 a.u. and r\textsubscript{2}
= 5.0 a.u.. Similarly to the approximation of the rotationally independent
terms, the bivariate surfaces $a_{ij}\left(\alpha,\tau\right)$ are
smooth and, thus, easily approximated by bivariate Chebyshev polynomials.
The operational parameters for the approximation of the $f\left(\theta,\varphi\right)$
and $a_{ij}\left(\alpha,\tau\right)$ surfaces are given in Table
\ref{tab:Operational parameters for the different levels of approximation tested}.
It was opted  to approximate the $a_{ij}\left(\alpha,\tau\right)$
and $f\left(\theta,\varphi\right)$ surfaces with the Chebyshev polynomials
of the same order.

\begin{table*}
\caption{\label{tab:Operational parameters for the different levels of approximation tested}Operational
parameters for the different levels of approximation tested}

\begin{tabular}{|c|c|c|c|c|c|c|c|}
\hline 
 & \multicolumn{7}{c|}{Functions of $g_{n}^{rot}\left(\overline{PQ}_{x}^{n}\right)$}\tabularnewline
\hline 
\hline 
Model & Order{*} & \multicolumn{3}{c|}{Threshold } & \multicolumn{3}{c|}{Order of Chebyshev polynomial for $b_{kl}^{ij}$}\tabularnewline
\hline 
 &  & $\varepsilon_{1}$ & $\varepsilon_{2}$ & $\varepsilon_{3}$ & Fine & Medium & Coarse\tabularnewline
\hline 
1  & 10 & 1.0 & 1.0E-02 & 1.0E-06 & 8 & 6 & 4\tabularnewline
\hline 
2  & 10 & 1.0E-01 & 1.0E-03 & 1.0E-07 & 8 & 6 & 4\tabularnewline
\hline 
 & \multicolumn{7}{c|}{Function of $g_{n}^{rot}\left(\overline{PA}_{x}\overline{PB}_{x}\right)$}\tabularnewline
\hline 
 &  & \multicolumn{3}{c|}{Threshold } & \multicolumn{3}{c|}{Order of Chebyshev polynomial for $b_{kl}^{ij}$}\tabularnewline
\hline 
 &  & $\varepsilon_{1}$ & $\varepsilon_{2}$ & $\varepsilon_{3}$ & Fine & Medium & Coarse\tabularnewline
\hline 
3 & 10 & 1.0E-01 & 1.0E-03 & 1.0E-07 & 8 & 6 & 4\tabularnewline
\hline 
4 & 11 & 1.0E-01 & 1.0E-03 & 1.0E-07 & 8 & 6 & 4\tabularnewline
\hline 
5 & 11 & 1.0E-03 & 1.0E-06 & 1.0E-09 & 10 & 8 & 6\tabularnewline
\hline 
\multicolumn{8}{|c|}{{*}Order of Chebyshev polynomial for $f\left(\theta,\varphi\right)$
and $a_{ij}\left(\alpha,\tau\right)$}\tabularnewline
\hline 
\end{tabular}
\end{table*}

The final step in the approximation of the rotationally dependent
functions is the fitting of the coefficients $b_{kl}^{ij}$. In Figure
6, the surfaces corresponding to $b_{1}^{1}$ and $b_{1}^{66}$ for
$g_{4}^{rot}\left(\overline{PQ}_{x}^{4}\right)$ are shown as a function
of the remaining coordinates r\textsubscript{1} and r\textsubscript{2}.
The dependency of the $b_{kl}^{ij}$ coefficients is considerably
simpler with the function being monotonically increasing in r\textsubscript{2}
(i.e. for fixed values of r\textsubscript{1}). Two important simplifications
can be introduced in the approximation of the $b_{kl}^{ij}$. First,
the order of the Chebyshev polynomials can be reduced since the surfaces
are easy to approximate, and second, many of the coefficients can
be eliminated since their contribution to $a_{ij}\left(\alpha,\tau\right)$
is negligible. Three parameters $\varepsilon_{i}$ are used to define
three regions of different approximating accuracy. In practice, the
maximum absolute value of $b_{kl}^{ij}\left(r_{1},r_{2}\right)$ is
compared with each parameter $\varepsilon_{i}$. If $\max|b_{kl}^{ij}\left(r_{1},r_{2}\right)|\ge\varepsilon_{1}$
the surface is approximated with the \textquotedblleft fine\textquotedblright{}
expansion of Chebyshev polynomials. \textquotedblleft Medium\textquotedblright{}
and \textquotedblleft coarse\textquotedblright{} expansions are used
when $\varepsilon_{1}>\max|b_{kl}^{ij}\left(r_{1},r_{2}\right)|\ge\varepsilon_{2}$
and $\varepsilon_{2}>\max|b_{kl}^{ij}\left(r_{1},r_{2}\right)|\ge\varepsilon_{3}$.
Surfaces for which $\max|b_{kl}^{ij}\left(r_{1},r_{2}\right)|<\varepsilon_{3}$
are discarded and the corresponding coefficients $c_{mn}^{ij,kl}$
are zero. The values of the $\varepsilon_{i}$ parameters and the
orders of the Chebyshev expansions for the \textquotedblleft fine\textquotedblright ,
\textquotedblleft medium\textquotedblright{} and \textquotedblleft coarse\textquotedblright{}
regions are also in Table \ref{tab:Operational parameters for the different levels of approximation tested}.
The quality of the different approximants of $f\left(\theta,\varphi\right)$,
$a_{ij}\left(\alpha,\tau\right)$, and $b_{kl}^{ij}\left(r_{1},r_{2}\right)$
is gauged in Table \ref{tab:Goodness-of-fit estimates and CPU times for terms for rotationally dependent terms},
where goodness-of \textendash fit results are presented and compared
for the $g^{rot}$ term. A distinction was made between the $g_{n}^{rot}\left(\overline{PQ}_{x}^{n}\right)$
and $g_{n}^{rot}\left(\overline{PA}_{x}\overline{PB}_{x}\right)$
terms because of their very different contributions to the total computed
ERI. Because of the small contributions to the computed ERI the $\overline{PQ}_{x}^{n}$
terms make, the long expansions of order 10 that were used in the
approximation of the $f\left(\theta,\varphi\right)$ and $a_{ij}\left(\alpha,\tau\right)$
surfaces were probably overkill. The RMSE and R\textsuperscript{2}
values are nevertheless excellent, indicating overall accurate approximations.
Special care was placed on the approximation of $g_{n}^{rot}\left(\overline{PA}_{x}\overline{PB}_{x}\right)$
because of the very significant partial contribution to the total
ERI. Thus, the rotational invariance of the approximated ERI greatly
depends on $g_{n}^{rot}\left(\overline{PA}_{x}\overline{PB}_{x}\right)$.
From Table \ref{tab:Goodness-of-fit estimates and CPU times for terms for rotationally dependent terms},
the most stringent model 5 seems necessary to achieve an excellent
accuracy in the approximation. It is important to stretch that approximation
with bivariate Chebyshev polynomials provides a way to systematically
improve the quality of the approximation. For example, the RMSE decreased
an order of magnitude on going from models 3 and 4 to model 5.

\begin{table*}
\caption{\label{tab:Goodness-of-fit estimates and CPU times for terms for rotationally dependent terms}Goodness-of-fit
estimates and CPU times for terms $g_{n}^{rot}\left(\overline{PQ}_{x}^{n}\right),n=1,2,3,4$
and $g_{n}^{rot}\left(\overline{PA}_{x}\overline{PB}_{x}\right)$}

\begin{tabular}{|c|c|c|c|c|c|}
\hline 
\multicolumn{6}{|c|}{$g_{n}^{rot}\left(\overline{PQ}_{x}^{n}\right),n=1,2,3,4$ }\tabularnewline
\hline 
\hline 
 & $\overline{PQ}_{x}^{4}$ & $\overline{PQ}_{x}^{3}$ & $\overline{PQ}_{x}^{2}\left(l_{1}+l_{2}=0\right)$ & $\overline{PQ}_{x}$ & $\overline{PQ}_{x}^{2}\left(l_{1}+l_{2}=2\right)$\tabularnewline
\hline 
\multicolumn{6}{|c|}{Model 1}\tabularnewline
\hline 
RMSE  & 4.03E-06 & 7.81E-06 & 2.82E-05 & 3.28E-05 & 1.46E-05\tabularnewline
\hline 
R\textsuperscript{2}  & 0.99997 & 0.99821 & 0.99950 & 0.97363 & 0.99973\tabularnewline
\hline 
time (s) & 555.7\textpm 3.4 & 546.4\textpm 3.9 & 550.0\textpm 3.3 & 549.4\textpm 2.4 & 554.4\textpm 4.3\tabularnewline
\hline 
\multicolumn{6}{|c|}{Model 2}\tabularnewline
\hline 
RMSE  & 7.94E-07 & 1.18E-06 & 4.63E-06 & 4.99E-06 & 3.29E-06\tabularnewline
\hline 
R\textsuperscript{2}  & > 0.99999  & 0.99996 & 0.99999 & 0.99943 & 0.99999\tabularnewline
\hline 
time (s) & 557.3\textpm 6.9 & 549.4\textpm 2.4 & 555.0\textpm 4.5 & 562.3\textpm 3.7 & 564.1\textpm 7.8\tabularnewline
\hline 
 &  &  & $g_{n}^{rot}\left(\overline{PA}_{x}\overline{PB}_{x}\right)$ &  & \tabularnewline
\hline 
 &  &  & RMSE & R\textsuperscript{2}  & time (s)\tabularnewline
\hline 
\multicolumn{3}{|c|}{Model 3} & 6.07E-04 & 0.99929 & 557.4\textpm 6.2\tabularnewline
\hline 
\multicolumn{3}{|c|}{Model 4} & 6.06E-04 & 0.99929 & 552.9\textpm 4.3\tabularnewline
\hline 
\multicolumn{3}{|c|}{Model 5} & 1.59E-05  & >\textcompwordmark{}> 0.99999 & 706.2\textpm 12.6\tabularnewline
\hline 
\end{tabular}
\end{table*}

\begin{figure}
\includegraphics[scale=0.4]{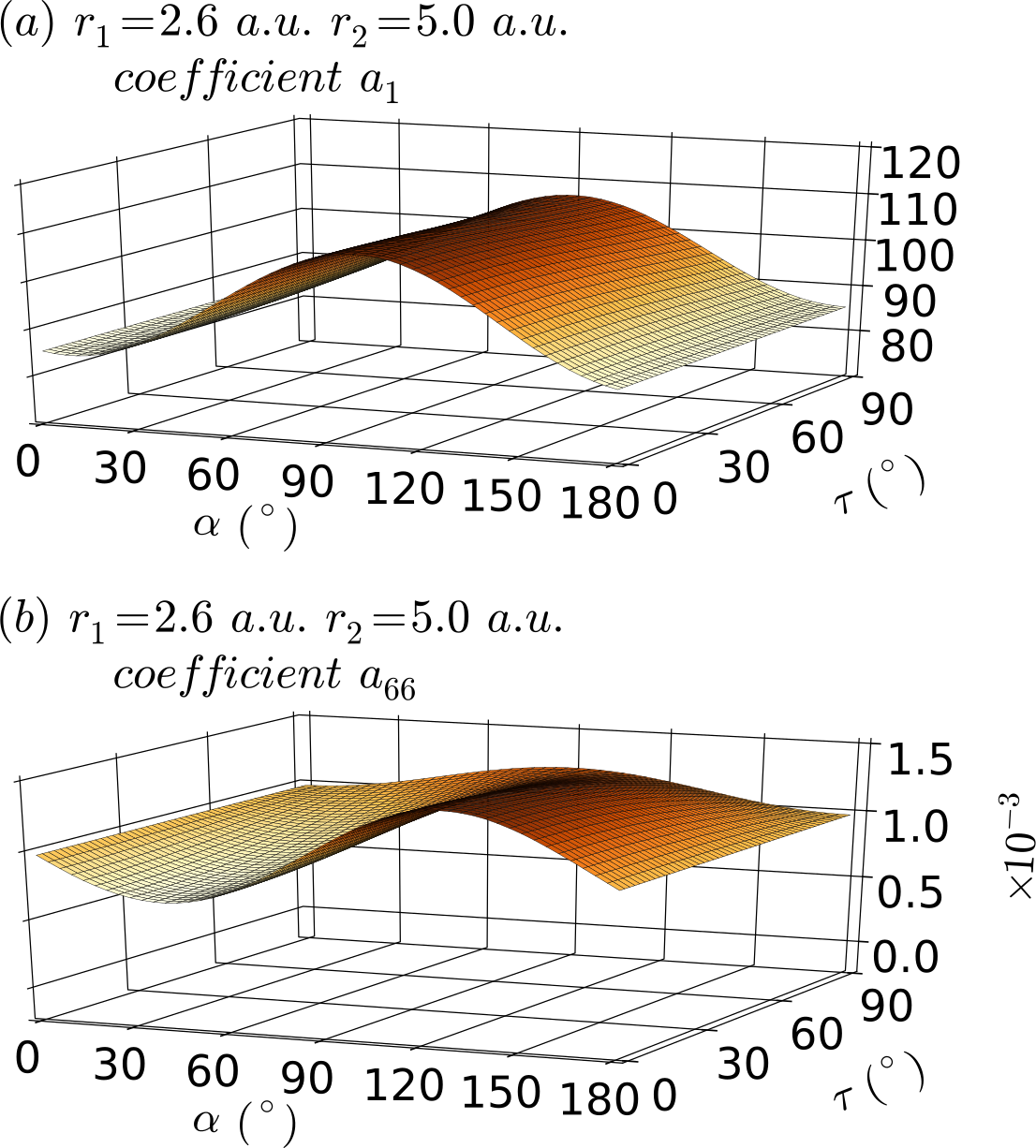}\caption{Illustration of dependency of the coefficients $a_{ij}\left(\alpha,\tau\right)$
of Equation 24a on the angles \textgreek{a} and \textgreek{t} for
fixed values of r\protect\textsubscript{1} = 2.60 a.u. and r\protect\textsubscript{2}
= 5.0 a.u. for $g_{4}^{rot}\left(\overline{PQ}_{x}^{4}\right)$. The
surfaces are smooth and have distinctive magnitudes that will be explored
to reduce the order of the polynomials used in the fittings.}
\end{figure}

\begin{figure}
\includegraphics[scale=0.35]{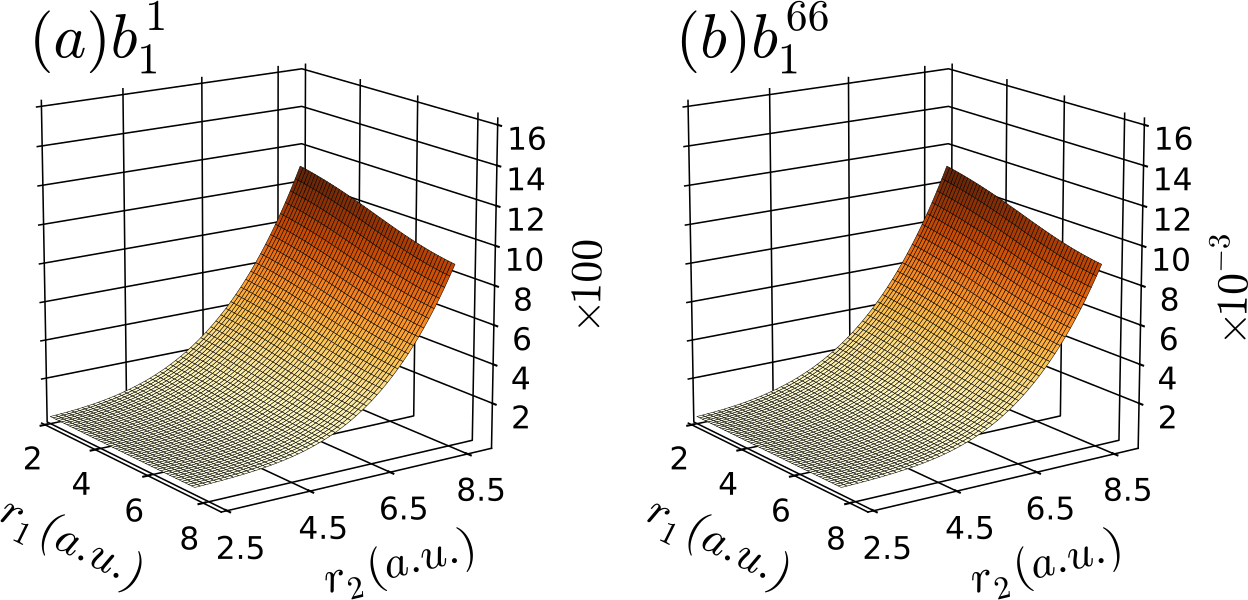}

\caption{Illustration of the dependency of the coefficients $b_{kl}^{ij}\left(r_{1},r_{2}\right)$
of Equation 24b on the distances r1 and r2 for $g_{4}^{rot}\left(\overline{PQ}_{x}^{4}\right)$.
The surface on the left (a) is for $b_{1}^{1}\left(r_{1},r_{2}\right)$
and the surface on the right (b) is for $b_{1}^{66}\left(r_{1},r_{2}\right)$.
Both surfaces are monotonically increasing in the r\protect\textsubscript{2}
direction, i.e. for fixed r\protect\textsubscript{1}. These surfaces
can be approximated with more compact Chebyshev polynomials, thus
reducing the overall computational cost.}
\end{figure}

\subsection{\label{sub:Adding-all-together 4C}Adding all together: assembly
of the computed $\mathbf{\left(p_{xA}p_{xB}|p_{xC}p_{xC}\right)}$
two-electron-repulsion integral }

The culmination of this work on the numerical approximation of ERIs
is the assembly of the calculated values of the$\left(p_{xA}p_{xB}|p_{xC}p_{xC}\right)$
ERIs from the different terms discussed above and the comparison with
the real analytical values. The contributors to the calculated ERI
are: the rotationally invariant term $G_{0}$, the rotationally dependent
terms $g_{n}^{rot}\left(\overline{PQ}_{x}^{n}\right)\cdot G_{n}$
and $g_{n}^{rot}\left(\overline{PA}_{x}\overline{PB}_{x}\right)\cdot G_{\overline{PA}_{x}\overline{PB}_{x}}$.
The approximation of $G_{0}$ was unique, using the highest order
expansion of this work. For the terms derived from $g_{n}^{rot}\left(\overline{PQ}_{x}^{n}\right)$,
two models were analyzed, but only model 2 was included in the assembly
of the final approximated ERI. The three models tested in the approximation
of $g_{n}^{rot}\left(\overline{PA}_{x}\overline{PB}_{x}\right)$ were
included in the computation of the approximated ERI. Goodness-of-fit
results for the total ERI are shown in Table \ref{tab:Goodness-of-fit estimates and total CPU timings for the approximated ERI}.
The results are the expected, and follow the pattern obtained for
$g_{n}^{rot}\left(\overline{PA}_{x}\overline{PB}_{x}\right)$ (Table
\ref{tab:Goodness-of-fit estimates and CPU times for terms for rotationally dependent terms}).
It is important to stretch not only the magnitude, but also the distribution
of the residuals. In Figure 7, the residuals for r\textsubscript{1}
= 2.6 a.u., r\textsubscript{2} = 5.0 a.u., \textgreek{a} = 60/145º
and \textgreek{t} = 30/60º are plotted as a function of $\theta$
and $\varphi$ for ($G_{0}$ + model 2 + model 4) and ($G_{0}$ +
model 2 + model 5). The biggest residual is roughly one order of magnitude
smaller with model 5 than with model 4, but importantly with model
5 the residuals are less than 1.0E-04 for most of the $\theta$ and
$\varphi$ domains, being higher in a very restricted area around
$\theta=0^{o}$ and $\varphi=90^{o}$. In contrast, with the less
accurate model 4 (also with the similar model 3) the residuals have
significantly higher values across the whole domains of $\theta$
and $\varphi$. The same pattern was found for other points and in
the future an exhaustive statistical study will be performed to determine
the validity of these anecdotal observations. The implications are
tremendous since it suggests the possibility of extending the areas
of extreme accuracy by redefining the limits of the domains where
the Chebyshev polynomials are defined. 

The timings of the calculation of the different rotationally dependent
terms are given in Table \ref{tab:Goodness-of-fit estimates and CPU times for terms for rotationally dependent terms}.
Despite the amazing results, with speedups of 4-5 orders of magnitude,
consideration of the timings required for the numerical approximation
of ERIs is secondary in this work. The calculations were performed
taking advantage of the highly structured grid points and are hardly
representative of real world scenarios. However, no specific optimizations
were developed and the Chebyshev polynomials were performed directly
using Eq. 19. In the future, specific optimizations will be introduced.
For example, the most costly calculation of the $b_{kl}^{ij}$ terms
according to Eq. 24c can be vectorized, introducing considerable speed
gains. The newly developed algorithm is inherently fast, requiring
only matrix-vector or matrix-matrix multiplications, operations that
are highly optimizd on multiple computer architectures, including
GPUs.

\begin{table}
\caption{\label{tab:Goodness-of-fit estimates and total CPU timings for the approximated ERI}Goodness-of-fit
estimates for the approximated $\left(p_{xA}p_{xB}|p_{xC}p_{xC}\right)$
ERI }

\begin{tabular}{|c|c|c|}
\hline 
 & RMSE & R\textsuperscript{2} \tabularnewline
\hline 
\hline 
$G_{0}$+ Model 2 + Model 3  & 6,07E-04 & 0,99929\tabularnewline
\hline 
$G_{0}$+ Model 2 + Model 4 & 6,06E-04  & 0,99929\tabularnewline
\hline 
$G_{0}$+ Model 2 + Model 5 & 1,59E-05  & >\textcompwordmark{}> 0,99999\tabularnewline
\hline 
\end{tabular}
\end{table}

\begin{figure*}
\includegraphics[scale=0.65]{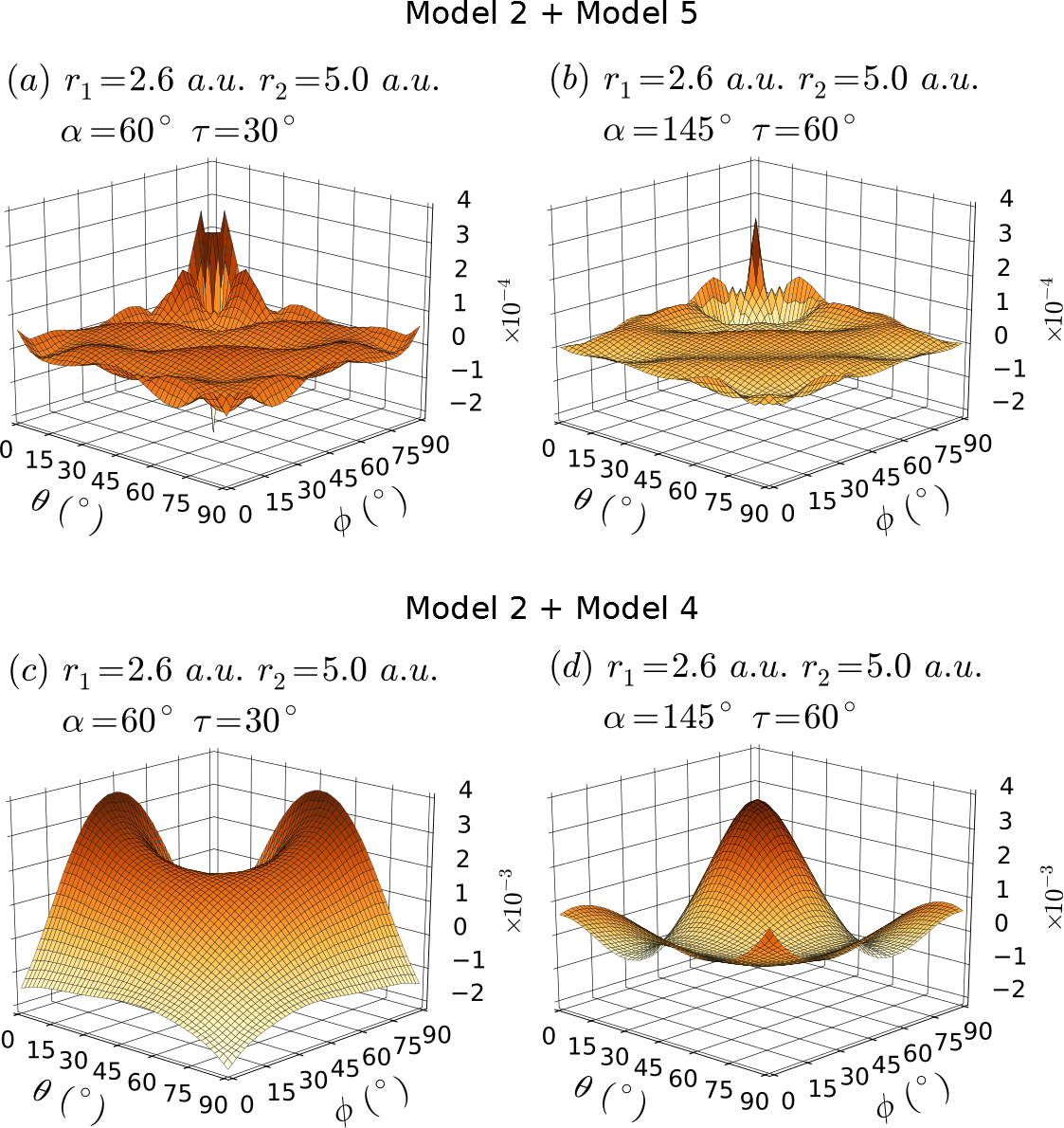}

\caption{Illustration of the residuals of the total computed ERI for two different
models of Table \ref{tab:Operational parameters for the different levels of approximation tested}:
model 2 + model 4 and model 2 + model 5. The rotationally dependent
terms $g_{n}^{rot}\left(\overline{PQ}_{x}^{n}\right)$ were only fitted
with the parameters of model 2. In contrast, the term $g_{n}^{rot}\left(\overline{PA}_{x}\overline{PB}_{x}\right)$
was fitted with both model 4 and model 5. The impact of model 5 on
the overall accuracy is impressive, with the residuals being one order
of magnitude smaller than with model 4 and importantly the biggest
absolute residuals very localized near $\theta=0^{o}$ and $\varphi=90^{o}$}
\end{figure*}

\section{\label{sec:Conclusions-and-future-prospects 5}Conclusions and future
prospects }

The work presented in this publication is the first step of a large
effort to develop novel tight binding computational methodologies
that are able to study large, complex systems. In the path to faster
and more generic computational quantum methods, three aspects are
the most significant: 1) computation of ERIs, theoretically an O(N\textsuperscript{4})
process, 2) diagonalization, itself an O(N\textsuperscript{3}) process,
and 3) the SCF iterations. The focus of this work was on the efficient
and accurate computation of ERIs. The approach consisted in using
multivariate approximation techniques to reproduce pre-computed target
ERI data. It is a proof-of-concept work aimed at demonstrating the
feasibility of such approximations. To my best knowledge, this was
the first time that such techniques were published. The test system
was the three-center ERI $\left(p_{xA}p_{xB}|p_{xC}p_{xC}\right)$
with all atoms being carbons. Having all atoms the same, introduces
important symmetry relations that help to simplify the amount of data
required for the approximations. In the initial phase of development,
when multiple calculations wre need in order to generate adequate
target data, it was important to keep the number of calculations to
a minimum. The same holds for the target basis set. The small STO-6G
was used because it allows efficient calculation of the many ERIs
required as target data. The methodology for the numerical approximation
consisted in decomposing a six-variable problem and a three-variable
problem into three bivariate problems and one univariate plus one
bivariate problem, respectively. The chosen approximating functions
were bivariate Chebyshev polynomials and a univariate polynomial or
order 10. The assumption was that for each sequential variable reduction,
the approximating coefficients yield a continuous function that can
be approximated by another set of polynomial approximants. The feasibility
of the methodology relies on assuming that the approximating coefficients
of a certain layer determine a continuous surface that can be fitted
in the next layer of approximations using the same technique. Although
no mathematical justification was attempted, it was indeed verified
that all surfaces and the single curve are continuous and could be
easily approximated. It is important to remember that the novel methodology
to approximate ERIs is not general and is not intended to replace
existing basis sets. 

The results are excellent with very small errors. In plots of residuals
for two specific points, it was found that for most of the approximating
domains the absolute error was significantly less than 1.0E-04. This
means that the approximating methodology is able to maintain the rotational
invariance of the computed integrals. Importantly, the new approach
does not depend on the size of the contractions of the basis set.
Although it was not a priority of this work, and no special attempts
were made to optimize the speed of the numerical approximations, the
methodology is very fast. Normalizing the CPU time to 1 core of the
computer system used in the work (AMD 8350), computation of the total
number of ERIs used in the fittings, more than 937 million, could
be completed in minutes. The same calculation of the analytical ERIs
on the same hardware would require approximately three months. 

The first major development in the future will be the creation of
a library of approximated ERIs, starting with the most common elements
in Biology: H, C, N, O, S, Na, K, Cl, Fe, Zn, Cu, and Ni. Appropriated
long single- and double-\textgreek{z} basis sets will also need to
be developed for these elements. A new tight-binding approach will
be implemented. It will incorporate the new methodology for fast diagonalizations
that was recently developed and will be linked to the library of approximated
ERIs. In a future publication, the new efficient diagonalization techniques
will be described. 

In conclusion, this work opens new perspectives to the future of computational
chemistry in general, for example to molecular simulations of large,
complex systems. The efficient computation of ERIs eliminates a significant
barrier to the generalization of computational quantum methods to
large systems. The combination of the methods for fast evaluation
of ERIs with novel approaches to diagonalize very large matrices will
allow development of specialized quantum based methodologies that
will be simultaneously fast and accurate.
\begin{acknowledgments}
P.E.M.L. wishes to thank M.M.G and J.D.N for support and M.S.L. for
reading the manuscript. \end{acknowledgments}

\end{document}